\documentclass[a4paper,fleqn,usenatbib]{mnras}

\usepackage[T1]{fontenc}
\usepackage{ae,aecompl}
\usepackage{booktabs}

\usepackage{graphicx}	
\usepackage{amsmath}	
\usepackage{amssymb}	

\title[Investigation of NGC~2808 with the UVIT]{Investigation of the Globular Cluster NGC~2808 with the Ultra-Violet Imaging Telescope}

\author[Rashi J. et al.]{
Rashi Jain$^{1}$, S. Vig$^{1}$\thanks{E-mail: sarita@iist.ac.in } and
S. K. Ghosh$^{2}$
\\
$^{1}$Indian Institute of Space Science \& Technology, Thiruvananthapuram 695547, Kerala, INDIA\\
$^{2}$Tata Institute of Fundamental Research,  Colaba, Mumbai 400005, INDIA\\
}

\date{Accepted XXX. Received YYY; in original form ZZZ}

\pubyear{2018}

\begin{document}
\label{firstpage}
\pagerange{\pageref{firstpage}--\pageref{lastpage}}
\maketitle

\begin{abstract}
Globular clusters represent stellar laboratories where observations can be used to validate models of stellar evolution. In this study, we put forth new ultraviolet (UV) photometric results of stars in the Galactic globular cluster NGC~2808. NGC~2808 is known to host multiple stellar populations that include at least four distinct groups of horizontal Branch (HB) stars. We have observed this cluster with the AstroSat-UltraViolet Imaging Telescope in two far-UV (FUV) and five near-UV (NUV) filters, respectively. These UV filters enable the identification of HB populations of stars. The results from four NUV filters exhibit bimodal distributions in magnitude histograms. The nature of bimodality has been investigated on the basis of distinct stellar types contributing to those bands. The color-magnitude diagrams constructed using FUV and NUV filters enable the location of hot stellar populations, viz. stars belonging to Red HB (RHB), Blue HB, Extreme HB, Blue Hook branch and post-Asymptotic Giant Branch. Prominent gaps are observed in the UV color-magnitude diagrams. We report for the first time, a photometric gap in a NUV color-magnitude diagram, that segregates the RHB population of this cluster into two groups, that are likely to be associated with distinct generations of stars. We also investigate the spatial density distributions of various groups of stars in the cluster and comment on the proposed formation models of multiple populations.
\end{abstract}

\begin{keywords}
\textit{(Galaxy:)} globular cluster: individual: NGC~2808 -- stars: horizontal branch -- \textit{(stars:)} Hertzsprung-Russell and color-magnitude diagrams -- techniques: photometric
\end{keywords}



\section{Introduction}

Stars in Globular Clusters (GCs), being the oldest population in the Galaxy, form an ideal test bed for
stellar evolution theories owing to their sheer numbers, metallicities, assorted stellar populations, and visibility due to negligible interstellar medium in the halo. In addition, having a statistical advantage of millions of stars as members, most of the stars are well resolved
for a detailed scrutiny leading to a sampling of even the short-lived populations \citep{1993ASPC...50....1V}. Although significant advances have been made in our understanding regarding the evolution of stars, ambiguities and gaps remain \citep{2001PASP..113.1162M}. 
Unlike the optical wavebands where a large body of work on GCs has been carried out, ultraviolet (UV) studies of
globular clusters have been relatively sparse \citep{2012AJ....144..126D,2005ApJ...625..156D}. Certain evolved populations are brighter in UV as compared to optical, due to various effects that include significant mass loss and metallicity. The main UV contributors in GCs are the horizontal branch (HB) stars, the Asymptotic Giant Branch (AGB) population, the blue stragglers, and the white dwarf stars \citep{1972A&A....18..390Z,1983PASP...95..256H}. 

The HB stars are central core helium burning low mass stars whose effective temperatures are expected to reach upto $\sim45,000$~K \citep{1995ApJ...446..228C}. They produce copious amounts of UV flux during this core He burning (CHeB) stage. The horizontal branch itself is not comprised of a simple stellar population and it is an aggregate of groups of stars separated by gaps in the color-magnitude diagrams, i.e the red horizontal branch, blue horizontal branch and extreme horizontal branch \citep{2011MNRAS.410..694D,1998ApJ...494..265C,1998ApJ...500..311F}. Among the HB group of stars, the red horizontal branch (RHB) stars are the cooler stars which lie towards  the red end of the HB. Their effective temperatures are expected to be in the range $\sim$5,000 - 6,200 K \citep{2013EPJWC..4302003A}. These have also been referred to as Red Clump stars \citep[eg. ][]{2016ApJ...822...44B,2008MNRAS.390..693D}. In accordance with the definition of \citet{2009Ap&SS.320..261C} and following \citet{2014MNRAS.437.1609M}, we address these stars as RHB in the present work. These stars are cooler than the RR Lyrae instability strip. The blue HB (BHB) stars lie on the hotter bank of the instability strip. In addition, there is a group of stars which are very luminous in far-UV, and these are believed to be in their post core
helium burning stage or post AGB (PAGB) phase. 

It is still not understood how various parameters affect the HB morphology. One of the parameters is the mass loss during the RGB phase. The HB stars have a core mass of $\sim0.5$~M$_{\odot}$ surrounded by a hydrogen rich envelope \citep{1970ApJ...161..587I,2010MmSAI..81..838M}. The mass of hydrogen envelope that surrounds the core could be related to the spread of stars along the blue HB, and those with thinner envelopes are expected to reside towards farther end of the blue HB. If the mass of H rich envelope exceeds $0.02$~$M_{\odot}$, the H shell burning is expected to remain active during the HB lifetime and later as it evolves towards the AGB phase, subsequent to the exhaustion of helium in it's core. These represent the blue horizontal branch (BHB) stars. On the other hand, if the mass of H envelope is less than $0.02 M_{\odot}$, the H shell burning cannot be sustained and the star directly evolves towards the white dwarf sequence. These hot but faint stars which are lower in mass are called extreme horizontal branch (EHB) stars. They do not ascend along AGB but proceed towards becoming white dwarfs. They have effective temperatures ranging between 20,000~K and 31,500 K \citep{1997A&A...319..109M}. The transition from BHB to EHB stars takes place at the fainter end of the blue tail of a globular cluster in UV-optical CMDs \citep{2010MmSAI..81..838M,2010ASSP...15...51F}. Unlike the RHB and BHB stars which have undergone helium flash at the tip of the Red Giant Branch (TRGB), the EHB stars are those that have suffered heavier mass-loss and are likely to experience helium flash anywhere between the TRGB to the top of white dwarf cooling curve. Stars that suffer from even higher mass loss, undergo a helium flash at even later stages of their descent along the white dwarf cooling curve. These stars are known as `late hot flashers' or blue hook stars \citep{2001ApJ...562..368B}. Blue Hook (BHk) stars constitute a rare group of stars that have been found in few GCs \citep{2002A&A...395...37M}. Metallicity is another parameter that is believed to play a key role in determining the morphology of HB. Metal poor clusters tend to have a HB that is extended farther out towards the bluer end, with conspicuous EHB and blue hook population \citep{2011MNRAS.410..694D}. 

The presence of RHB and BHB is found in many GCs, although the fractional populations and HB morphology may vary by degrees. High resolution spectroscopy of few GCs have revealed metallic abundances and chemical inhomogeneities that are expressed as multiple main sequences, broadened subgiant branches, and split red giant branches, that are referred to as multiple stellar populations \citep{2004MSAIS...5..105B,2012ApJ...760...39P,2010AJ....140..631B,2016MmSAI..87..303M}. The puzzling chemical compositions analysed through extensive spectral and photometric investigations, and have been ascribed to mechanisms such as multiple star-formation bursts, \citep{2007A&A...464..967C}, rotation effects \citep{1977asse.conf..151R}, pollution by AGB stars \citep{2004ARA&A..42..385G}, and unusually high mass-loss \citep{2001ApJ...562..368B}.

The UV arena provides a unique opportunity to investigate the hot stellar populations of globular clusters, thus enabling a glimpse into processes dominating in the later evolutionary stages of stars. The UV studies ($\lambda\lesssim300$~nm) are difficult to accomplish owing to atmospheric opacity, and our understanding of globular clusters has benefitted immensely from the few UV space missions that have been achieved, such as the Hubble Space Telescope \citep[HST;][]{2013ApJ...770...45D}, the Galaxy Evolution Explorer \citep[GALEX;] []{2012AJ....143..121S}, the International Ultraviolet explorer \citep[IUE;][]{1982AdSpR...2..119C}, and the SWIFT-Ultraviolet/Optical Telescope \citep[UVOT;][]{2007AAS...211.7305H}, to name a few. The latest observatory to join the group is AstroSat which houses the Ultraviolet Imaging Telescope (UVIT). UVIT, with its multiple far- and near-UV filters provides a novel opportunity to analyse the hot populations of globular clusters. The combination of large field-of-view ($\sim28'$) and moderately high angular resolution ($1.2''$) facilitates the imaging and examination of intermediate and outer radial regions of globular clusters in exceptional detail. Towards this end, we have investigated NGC~2808, an old globular cluster with intermediate metallicity that possesses an extended horizontal branch and hosts multiple stellar populations.

\begin{figure}

\includegraphics[width=1\linewidth]{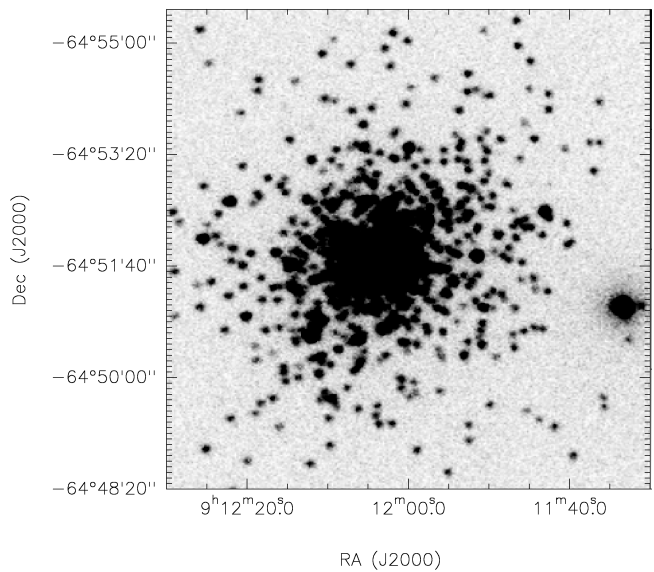}    

\caption{The Galex FUV broadband image of the globular cluster NGC~2808. }
\label{fig:gal}
\end{figure}

NGC~2808 is an old Galactic globular cluster  (age~$10.9\pm0.7$ Gyr) with a distance modulus $(m-M)_{V}=15.79$ \citep{2000A&A...363..159B}, and metallicity [Fe/H]$=-1.24$ \citep{1997A&AS..121...95C}. This GC displays a multimodal HB star distribution \citep{1974ApJ...192L.161H}. The HB exhibits three distinct gaps in the CMDs with an anomalous blue tail extending considerably towards very high temperatures. The blue HB of the cluster is found to have clumpy distribution with two narrow gaps \citep{1997ApJ...480L..35S} that bifurcate the HB stars into four distinct populations. These represent the RHB, BHB, EHB and BHk contents of the cluster. \citet{2011MNRAS.410..694D} correlated these branches of HB to the multiple populations of the main-sequence in the cluster, reported by \citet{2007ApJ...661L..53P}. The BHB stars were correlated to models with high helium content, Y=0.3 (progeny of normal-MS population), while RHB stars were matched by models with Y=0.24 (Red-MS) and EHB stars were mapped to Blue-MS (Y=0.4) by \citet{2008MNRAS.390..693D}. It is now established that NGC~2808 hosts at least five populations of stars \citep{{2015ApJ...808...51M},{2015ApJ...810..148C}}. An elaborate study has been carried out by \cite{2011A&A...528A.127M} to identify HB stars in NGC~2808 using spectroscopy. In addition, \cite{2011A&A...534A.123G} spectroscopically classified HB stars of NGC~2808 using Na-O anticorrelation. The UV wavebands provide a singular platform to testify the origin of a clumpy distribution of hot stars. Using GALEX, \cite{2012AJ....143..121S} investigated a number of GCs including NGC~2808 and distinguished Post-AGB and Post (early) AGB stars in these clusters using CMDs. The GALEX field of NGC~2808 in FUV is shown in Fig.~\ref{fig:gal}. In the current study, we investigate the hot population of the cluster in the ultraviolet realm of UVIT. 


In Sect. \ref{sec:obs}, we present the UV observations and procedures adopted for data reduction and analysis. The results are described in Sect. \ref{sec:result} and discussed in Sect. \ref{sec:discussion}. The article concludes with a brief summary in Sect. \ref{sec:con}.

\section{Observations and Data Reduction}
\label{sec:obs}
\subsection{AstroSat - UVIT Observations and pipeline reduction}
The UVIT on-board the AstroSat has been used to image the globular cluster NGC~2808. AstroSat is India's first space based multiwavelength mission \citep{2004cosp...35.1420A} that observes simultaneously in UV, visible and X-rays. UVIT has a twin telescope system with a diameter of 38 cm, in Ritchey-Chretien configuration. One of the telescopes is dedicated for observations in FUV and the second telescope observes in NUV and visible channels.  UVIT is primarily an imaging instrument and covers a circular field of view of $\sim 28'$. The spatial resolution achieved by UVIT in FUV and NUV is $< 1.5''$, while it is  $< 2.2''$  in the visible channel. Each channel (FUV, NUV and visible) constitutes of a number of selectable filters with smaller passbands. Further details about the UVIT and the instrument performance  can be found in \cite{2017AJ....154..128T}. 

NGC~2808 was observed by UVIT on April 4, 2017, in two FUV, and five NUV filters. The data from simultaneous imaging in a neutral density visible filter were used for aspect reconstruction during the post-observation data processing stage. The filters used during the observations were F154W and F169M in FUV, and N242W, N219M, N245M, N263M and N279N in NUV. The details of the filters are listed in Table~\ref{tab:Filters}. The exposure times of observations through various filters are also listed in the table. The Level-1 data products made available by the Indian Space Science Data Center (ISSDC/ISRO) were processed using the UVIT Level-2 Pipeline (UL2P) version V5.6. The final products generated by the UL2P include FITS images of the sky. The images cover the field of view with a pixel size $0.41'' \times 0.41''$. These images have been used for the present study. The pipeline corrects for various instrumental effects such as spacecraft drifts, jitters, thermal effects and other corrections required for astronomical imaging. The final output comprises of an image through each (NUV/FUV) filter along with the corresponding uncertainty image. The intensity unit of the images are in counts/second. The angular resolution achieved in the images is $\sim1.2''$, superior to that of GALEX ($\sim$6$''$). 


\begin{table*}
\centering
\caption{Details of UVIT filters and observations of the globular cluster NGC~2808.} 
\label{tab:Filters}
\begin{tabular}{lccccccccc}
\hline
\toprule
Band & Filter & Alternate & $\lambda_{mean}$ &$\delta\lambda $ & ZP mag & Exposure Time & Sensitivity Limit& Number of  & Number of \\
 & & Name & (\AA) & (\AA)& & (sec) & (AB mag for SNR=3.5) & stars & Saturated stars\\
\hline
FUV & F154W & BaF2 & 1541 & 380 & 17.77$\pm$0.01 & 4172 & 24.034 & 905 & 1\\
FUV & F169M & Sapphire & 1608 & 290 & 17.45$\pm$0.01 & 3553& 23.548 & 855 & 1\\
NUV & N242W & Silica & 2418 & 785 & 19.810$\pm$0.002 & 403 & 23.542 & 1556 & 21\\
NUV & N219M & NUVB15 & 2196 & 270 & 16.59$\pm$0.02 & 3494 & 22.667 & 905 & 2 \\
NUV & N245M & NUVB13 & 2447 & 280 & 18.50$\pm$0.07 & 477 & 22.415 & 1379 & 5\\
NUV & N263M & NUVB4 & 2632 & 275 & 18.18$\pm$0.01 & 348 & 21.752 & 1275 & 10\\
NUV & N279N & NUVN4 & 2792 & 90 & 16.50$\pm$0.01 & 2639 & 22.155 & 1810 & 1\\
\hline
\end{tabular}
\end{table*}



\begin{figure*}
\begin{minipage}[b]{0.45\linewidth}
    \includegraphics[width=1\linewidth]{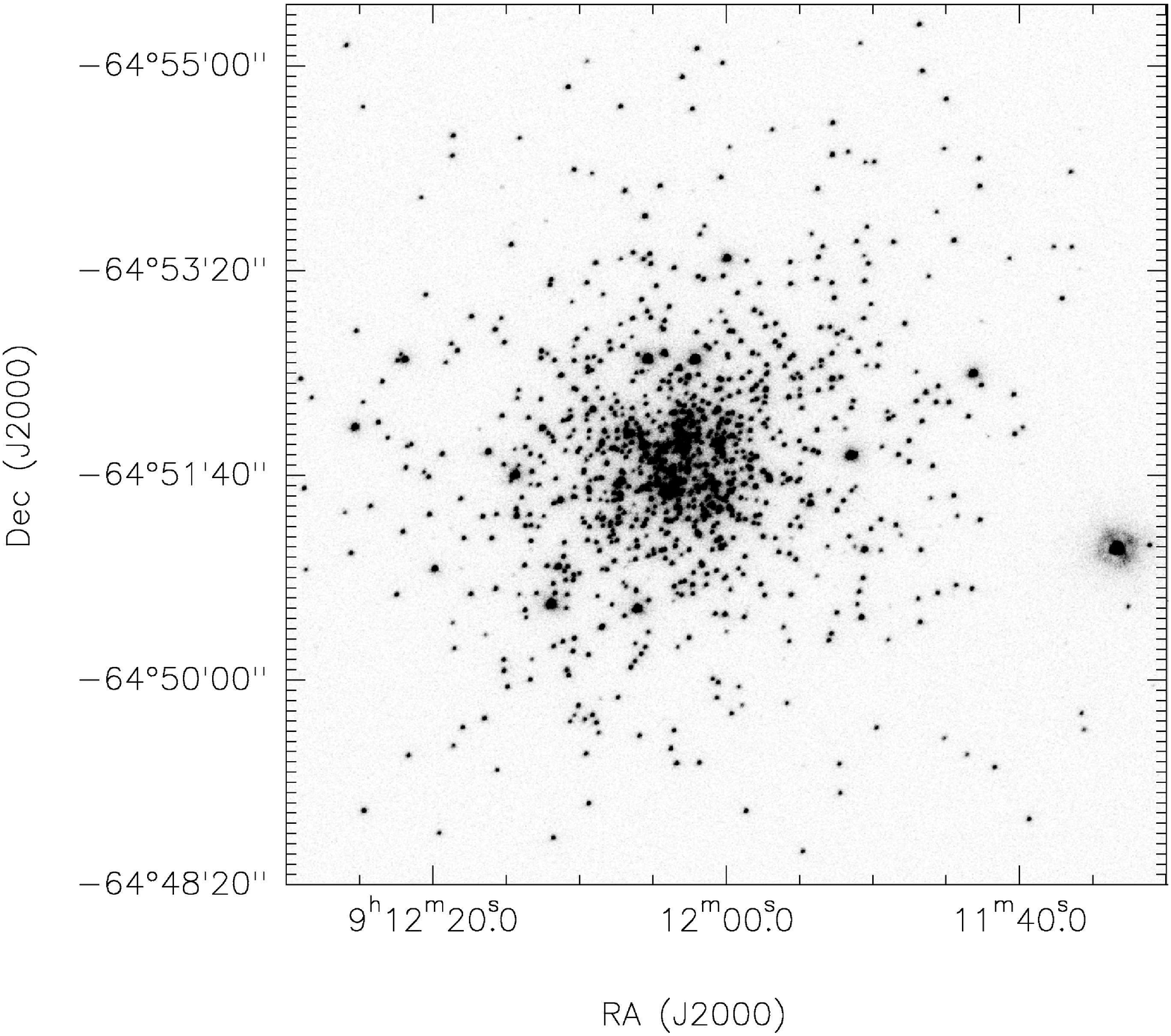}    
\end{minipage}
\begin{minipage}[b]{0.45\linewidth}
    \includegraphics[width=1\linewidth]{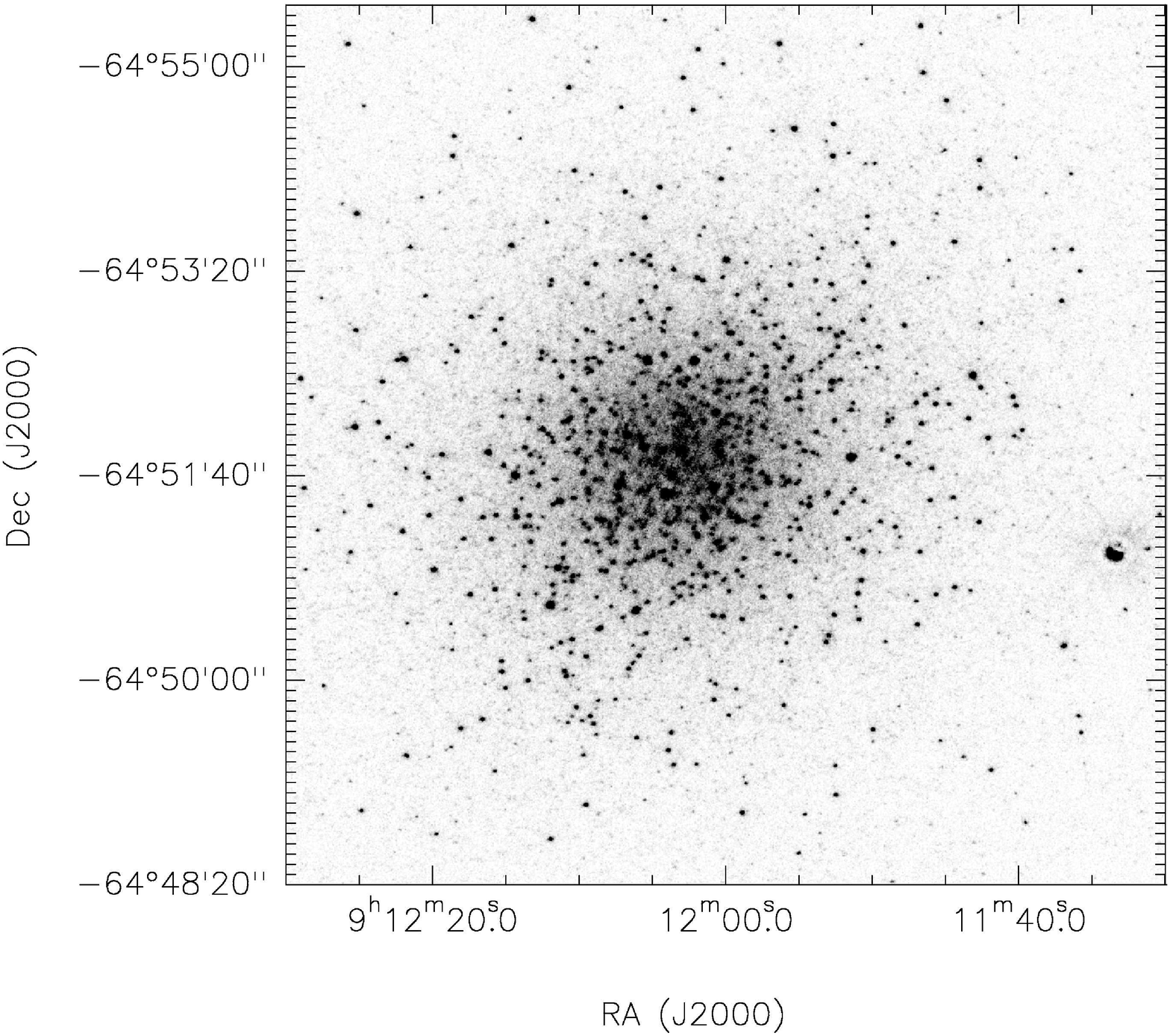}    
\end{minipage}
\caption{The UVIT images of the globular cluster NGC~2808. The far-UV F154W image is shown on the left panel, while the near-UV N242W image is displayed in the right panel.}
\label{fig:FUVG}
\end{figure*}


\subsection{Additional Data Reduction and Photometry}
Long (> 2000 sec) on-target observations of UVIT are generally split into exposures in multiple orbits (as UVIT can operate only during the `dark' part of each orbit when the spacecraft is in earth's shadow). The UL2P processes datsets from individual orbits to generate sky images from each of them as well as combined products from all orbits contributing to the on-target integration. The UL2P pipeline includes an Astrometry block at the very end of the processing sequence, which uses a set of brightest stars detected in the UV image and their coordinates are correlated with entries in the optical calalogue USNO-A2.0. In our case, the Astrometry block of UL2P was unable to refine the absolute aspect of the sky images due to the crowded field of NGC~2808. Accordingly, an additional reduction scheme was employed to improve the aspect. This scheme used IRAF task `CCMAP' by considering 10 bright stars from the USNO-A2.0 catalog distributed across the field-of-view. The positions of the sources are found to be uncertain upto $\sim1.0''$ after the corrections were applied.

The images of NGC~2808 through the FUV filter F154W and NUV filter N242W are displayed in Fig.~\ref{fig:FUVG}. A comparison between the images shows that NUV is substantially crowded across the field whereas FUV is relatively sparse, as expected. Thus, the stars in the inner central region were resolved in FUV, whereas in NUV the field is considerably crowded near the cluster core. 

\subsubsection{Photometry}
\label{sec:phot}
We performed photometry on the images in order to extract the magnitudes of stars in all the UV filters. For this, we used a combination of DAOPHOT and SExtractor (SE). We employed the DAOPHOT subroutine, DAOFIND, to locate the stars in the field and PHOT for the initial level aperture photometry. The results from PHOT are used to construct a PSF model using bright stars in the field and then the subroutine ALLSTAR was used to perform the PSF photometry. 
 
The central circular region (radius $< 20''$) in the NUV images is deliberately omitted from photometric analysis because of the inability of the software to resolve stars in this region. A number of faint stars were observed in the field in NUV, that were not detected by DAOPHOT. SE, on the other hand, performed better and was able to detect and carry out photometry on these stars. The integrated counts/sec corresponding to the brightness of a star through a filter was converted to the AB magntitude system after removing the sky background appropriately (which is minimal), taken from the neighbourhood. The AB magnitudes have been calculated using the following expression:

$$m_{AB}=m_{ZP}-2.5 \times log(CPS)$$

\begin{figure}
\includegraphics[width=1\linewidth]{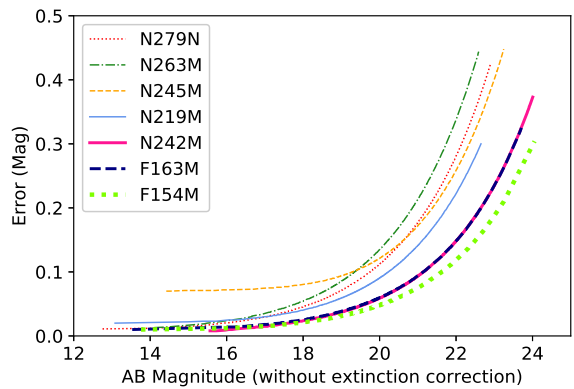}
\caption{The variation of error as a function of AB magnitude (without extinction correction) for the sources detected in the UVIT filters.}
\label{fig:err_mag}
\end{figure}

Here, $m_{ZP}$ represents the Zero-Point magnitude of the filter, taken from \citet{2017JApA...38...28T}, and $CPS$ is the integrated counts/sec due to the star.  The ZP magnitudes as well as their uncertainities are listed in Table~\ref{tab:Filters}. The error in AB magnitude for each UV filter has been estimated considering: (i) the uncertainties in CPS arising from Poisson noise of the photon-counting detector, and (ii) the error in the ZP magnitude from  calibration uncertainties. A plot showing the variation of errors as a function of magnitude is displayed in the Fig.~\ref{fig:err_mag}, each line representing one UV filter.


The bright stars in the field are likely to be affected by saturation. In particular, stars that are brighter than the Zero-point magnitude by 1.95 magnitude or more (in a given filter) are likely to be severely affected. We have corrected for saturation using the procedure described by \citet{2017JApA...38...28T}. The number of stars saturated in each filter is listed in Table~\ref{tab:Filters}. We find that the brightest star is heavily saturated in all the UV filters. This star is located at ($\alpha$, $\delta$)$_{\rm J2000}$ = $9^h 11^m 33.32^s$, $-64^\circ 51'03.30''$, which is $3.25'$ from the cluster centre. It was not possible to correct for saturation for this star and therefore, the given magnitudes of this stars represent lower limits to its brightness.

In addition to the above procedures, the field of NGC~2808 was also visually scrutinised through each filter in order to separate out closeby stars that were identified as a single star by DAOPHOT and SE. The magnitudes of these stars were estimated manually through aperture photometry. Aperture photometry was also carried out for few faint stars that were not picked by either of the softwares. Finally, the stars detected through each filter was compiled to construct a catalogue of UV stars. It is to be noted that the number of stars detected through a given filter depends on the exposure time and effective area of the filter. In order to have a magnitude limited sample through each filter, we considered those stars in the present study that were brighter than the magnitude corresponding to a signal-to-noise ratio of 3.5. The AB magnitudes corresponding to this sensitivity limit are given in Table~\ref{tab:Filters}. Note that these values represent native limits that have not been corrected for extinction. The total number of UV stars detected is 2253, of which 484 are common to all the UV filters. The number of stars detected through each UVIT filter is listed in Table~\ref{tab:Filters}. In the FUV, we detected more than 900 stars. Nearly double the number of stars were detected in few NUV filters.


\begin{table}
\caption{Extinction values through the UVIT filters corresponding to $A_{v}$=0.54 based on the extinction law given by \citet{1989ApJ...345..245C}.}
\centering
\label{tab:Red}
\begin{tabular}{l c c }
\hline 
\toprule
Filter & $\lambda_{mean}$ ($\nu$m) &$A_{\lambda} (mag)$ \\ 
\hline 

F154W & 1541 & 1.46 \\

F169M & 1608 & 1.43 \\

N242W & 2418 & 1.40\\

N219M & 2196 & 1.77 \\

N245M & 2447 & 1.36 \\ 

N263M & 2632 & 1.18 \\
N279N & 2792 & 1.09 \\
\bottomrule
\end{tabular}
\end{table}


\subsubsection{Extinction Correction}
Extinction correction of stellar magnitudes is important for ultraviolet wavebands. An accurate analysis requires the knowledge of foreground extinction of the cluster. The integrated color and spectral type of the cluster gives the value of selective extinction ($E_{B-V}$) to be 0.18 towards this cluster \citep{2000A&A...363..159B}. Adopting the ratio of total-to-selective extinction as $R_{v}$ = 3.1 \citep{1958AJ.....63..201W} for Milky Way, the extinction co-efficient in visible is $A_{V}=0.54$.
The $A_{V}$ is used to determine extinction co-efficients $A_{\lambda}$ for all passbands using the reddening relation of \cite{1989ApJ...345..245C}. Table \ref{tab:Red} summarizes the values of  $A_{\lambda}$ estimated by interpolation. 

\section{Results}
\label{sec:result}
Having carried out photometry on the images of NGC~2808 and constructed a catalog of UV stars, we proceed towards the understanding of hot stellar populations that contribute to UV. We begin with understanding the magnitude distribution of stars through various filters.

\subsection{UV Magnitude Distribution}

\begin{figure*}
\centering
\includegraphics[width=0.7\linewidth]{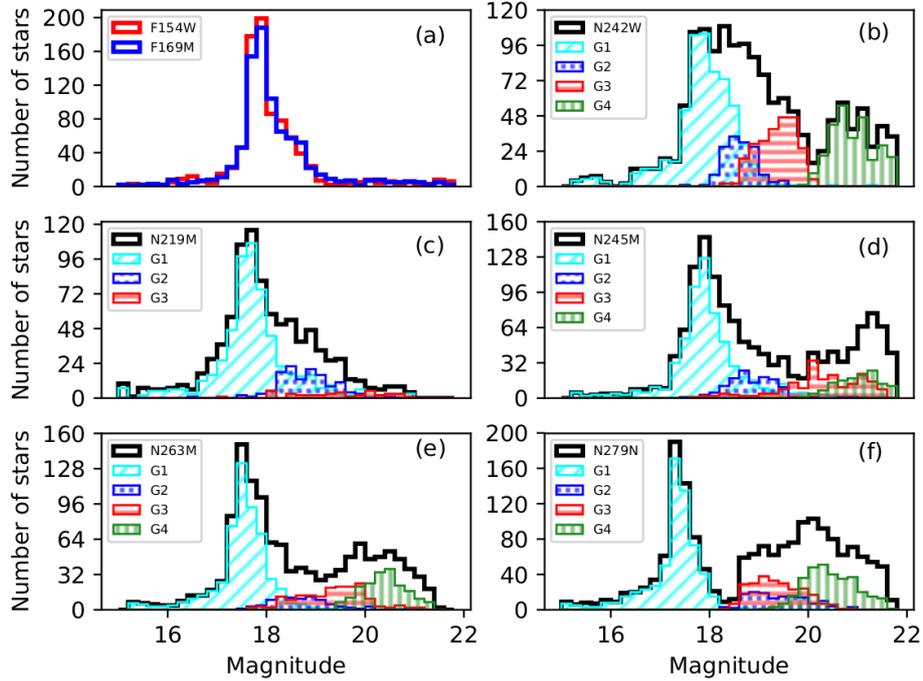}
\caption{The magnitude distribution of the cluster members of NGC~2808 through various AstroSat-UVIT filters. The total number of stars in each magnitude bin for NUV filters have been decomposed into various sub-groups of stars based on the classification from the CMD of N242W-N279N versus N242W (see Section \ref{sec:nuvnuv}). Note that $\sim60\%$ of total stars are detected in these two filters. The numbers of stars corresponding to each of the 4 sub-groups have been overplotted for each NUV filter. The histogram in the panel (a) shows the magnitude distribution of both the FUV filters, while panels (b), (c), (d), (e) and (f) exhibit the magnitude distributions through the NUV filters N242W, N219M, N245M, N263M, and N279N, respectively.}
\label{fig:mag_hist}
\end{figure*}

The magnitude distribution of the UV stars of NGC~2808 through the various filters of AstroSat are displayed in Fig.~\ref{fig:mag_hist}. We observe that the distributions through both the FUV filters are unimodal and quite similar. Most of the stars lie in the magnitude range $16-19$~mag, and the peak of the distribution is $\sim18$~mag. The magnitude distributions display a sharp decline from the peak towards the brighter magnitudes whereas the drop is more gradual towards the fainter side. We expect the FUV field to be dominated by the hottest sub-populations of HB stars which is likely to explain the lack of bimodality in the distribution of the FUV magnitudes. The similarity in the magnitude distributions is attributed to the fact that the cut-off wavelengths of the two FUV filters are quite similar ($1340$ \AA~for F154W, and $1420$ \AA~for F169M). This implies that the presence of any absorption lines in the cut-off interval, $1340-1420$ \AA~is unlikely to play an important role in the hot stellar populations sampled by FUV. The FUV spectroscopy of hot stars in this cluster by \citet{2012ApJ...748...85B} confirms this view. 

The NUV filters, on the other hand, exhibit unimodal and bimodal distributions. The filter N219M shows a single broad peak unlike the other NUV filters that display bimodality in their magnitude distributions. The presence of bimodal peaks primarily points towards the segregation of hot stellar populations in the respective filter, albeit in a complex way. The wide band NUV filter N242W displays two peaks, wherein the brighter (primary) distribution peaks at $\sim 18.5$~mag, has a larger number of stars, and is broad. The bimodal separation is observed at 20~mag in this filter. We note that sensitivity prevents the detection of cluster members beyond 22 mag, hence we refrain on commenting on the width of the secondary peak. The medium bandwidth filters include N219M, N245M and N263M. Among them, N219M which is the shortest waveband filter in NUV, displays  a single peaked distribution centered around 17.5~mag. Moreover, the distribution is perceived to be skewed with a gradual decline towards the fainter side of the peak, unlike the lowering towards the brighter edge.  The other two medium band filters, N245M and N263M  both display bimodality with the former showing a narrower secondary peak compared to the latter. N279N, the narrow band filter at the longest NUV waveband reveals a narrow primary distribution and broad secondary distribution. In the three NUV narrow and medium waveband filters displaying bimodality, we observe that the magnitude bin separating the primary and secondary distributions shifts towards the brighter side as the wavelength increases. This is likely to be related to the stellar population distribution and we discuss this in later sections.


\subsection{Identification of hot stellar populations}

NGC~2808 has been very well-studied at a multitude of wavelengths, spanning UV to near-infrared. Imaging as well as spectroscopy have led to an understanding of various stellar populations of this globular cluster. In this subsection, we elaborate on the identification of various hot stellar populations in our UV catalog from other works in literature. The purpose of the cross-identification is to understand and interpret the location of these populations in the color-magnitude space of the UV filters of AstroSat-UVIT. 

\begin{itemize}
 \item[-] \textit{Red Horizontal Branch and Blue Horizontal Branch stars}: \cite{2011A&A...534A.123G} spectroscopically identified 49 RHB stars and 36 BHB stars in NGC~2808. The identification is based on the anti-correlation between Na and O discerned in HB stars of globular clusters. The RHBs stars are characterised by Na poor and O rich content while the BHB stars have Na rich and O poor composition \citep{2011A&A...534A.123G,2009A&A...505..139C}. We find a match of 34 RHB and 35 BHB stars in the cluster as the rest are from the central region excluded from our analysis.
\item[-] \textit{Extreme Horizontal Branch stars}: \cite{2011A&A...528A.127M} identified 8 EHB stars  from the catalogue published by \cite{2003A&A...407..303M}. They employed a temperature criterion of the zero age HB model in their optical color-magnitude diagram to identify the association.  We have identified all the eight EHB sources in our catalogue.
\item[-] \textit{Blue-hook stars}: \cite{2004A&A...415..313M} performed ground based spectroscopy for stars in NGC~2808 which were identified as BHk Stars. All these stars are located beyond 2$'$ from center of cluster. All the 19 BHk stars from their sample have been matched in our catalogue.
\item[-] \textit{Post-AGB, Post-Early AGB and AGB-manque stars}: \cite{2012AJ....143..121S} have studied severals GCs in UV using GALEX, and published a list of PAGB, Post-early AGB and AGB-manque stars in these clusters. Towards NGC~2808, their list comprises 22 stars belonging to this UV-bright category. All these UV-bright stars have been identified in our catalogue.
\end{itemize}

We observe that there are overlaps between classifications among our cross-matched samples: 5 BHB stars are also classified as BHk stars, 2 stars are classified as BHB and PAGB, and 1 star is common in the catalogs of BHk and PAGB. 

\subsection{UVIT Color-Magnitude Diagrams}

In this section, we discuss the UV color-magnitude diagrams and analyse them based on the known stars of different stellar populations. We continue to refer to the term `color' inspite of the fact that responses of some filters overlap in wavelength. In other words, the `color' derived from overlapping filters cannot be directly used as a proxy of effective temperature, although they are expected to be related in an intricate way. In certain cases, where the filter responses are completely non-overlapping (e.g. colours corresponding to FUV - NUV  and N245M - N279N), they can directly probe the effective temperatures.

\subsubsection{FUV - NUV Color-Magnitude diagram}

\begin{figure}
\includegraphics[width=1\linewidth]{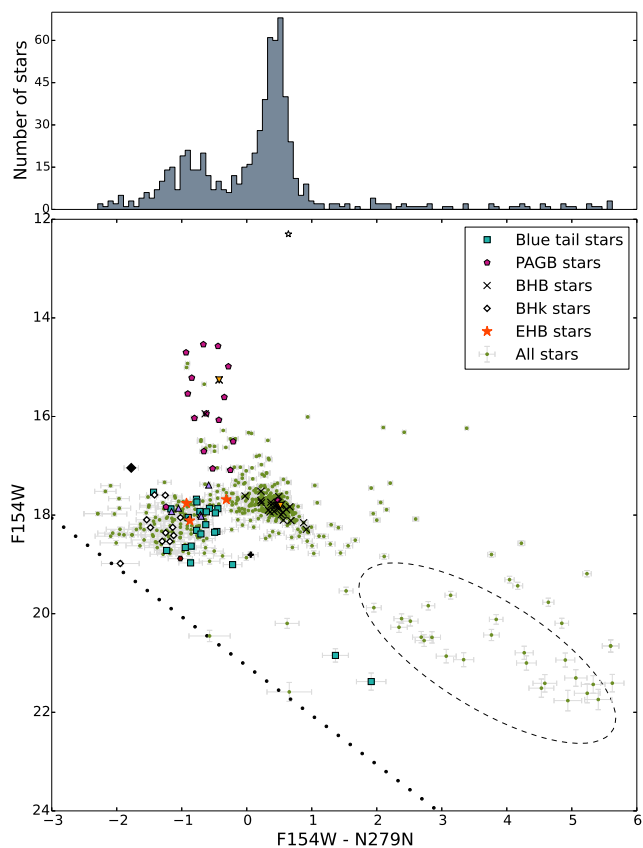}
\caption {The FUV - NUV color-magnitude diagram constructed using F154W - N279N versus F154W. The histogram on the color axis exhibits the color distribution of all the stars plotted here. Stars identified using other catalogues (see text for details) are marked using symbols shown on the top-right in the color-magnitude diagram.  The stars that have overlaps between catalogs are represented by black filled diamonds (BHk as well as PAGB), orange inverted triangles (BHB and PAGB), violet triangles (EHB as well as BHk), filled hexagon (BHk and blue tail), and `+' (BHB and blue tail). The dashed elliptical region encircles the region likely to be occupied by Blue straggler stars. The green filled circles are the other stars common to F154W and N279N filters which have not been identified earlier. The unfilled asterix represents the brightest star in the cluster, which is saturated in all bands. The sensitivity limit is translated to the dotted line shown in the figure.}
\label{fig:fuvnuvcmd}
\end{figure}

Considering that we have 2 FUV filters and 5 NUV filters, a large number of color - magnitude diagrams (CMDs) are feasible. However, in the present work, we consider only those diagrams where we discern a notable segregation of stars. Among the two FUV filters, we can select any one, as the number of stars and magnitude distributions are similar. We consider F154W for the present case. Among the NUV filters, we use N279N as we find that the colors associated with this filter exhibit a bimodal distribution unlike the other filters. The CMD of F154W - N279N versus N279N is displayed in Fig.~\ref{fig:fuvnuvcmd}. Also overplotted on this CMD are the BHB, EHB, BHk and PAGB stars identified from literature (Sect. 3.2). The known RHB stars were not detected in FUV and hence they are not present among the sample. We observe that the various stellar groups appear well separated. 
 

The BHB group of stars occupy the region associated within the color range -0.7 -- 1 in the FUV - NUV color-magnitude plane. A majority of the stars detected in these two filters belong to the BHB group, evident from the color histogram. The stars hotter then BHB stars occupy a region within the color range -2.5 -- -0.5 on the color axis. This group of hotter HB stars comprises of EHB and BHk stars of the cluster. The BHB group is separated from the group of EHB and BHk stars by a jump, which is reported to be a ubiquitous feature of the GCs hosting BHB and EHB stars \citep{2001ApJ...562..368B}. This jump, also known as Momany jump \citep[M-jump, ][]{2004A&A...420..605M} corresponds to a temperature $\sim$ 20,000K. In our filters, we associate this jump with a color corresponding to -0.7 mag. The BHks occupy the hotter end of the CMD while the EHB group is intermediate, between the BHk and BHB stars. There is a superluminous sub-population that occupies the magnitude range 17 - 14 in the FUV. These stars belong to the PAGB and Post-early AGB stages. The number of stars belonging to this group is smaller than the HB population, and they are sparsely distributed towards the brighter side of the magnitude range. There are a few stars that are subluminous in FUV and and are very red in the CMD, encircled within a dashed ellipse in  Fig.~\ref{fig:fuvnuvcmd}. It is possible that few of these belong to the blue straggler (BS) category \citep{2001ApJ...562..368B}. BS stars are more massive than the main sequence turn-off stars, and their presence is ascribed to mass transfer or collision \citep{2006MNRAS.373..361M,2001AAS...199.0607C}. The blue square symbols in the figure represent stars classified as blue-tail stars, defined later in Sect.~3.3.2.

\subsubsection{NUV - NUV Color-Magnitude Diagrams}
\label{sec:nuvnuv}

\begin{figure}
\includegraphics[width=1\linewidth]{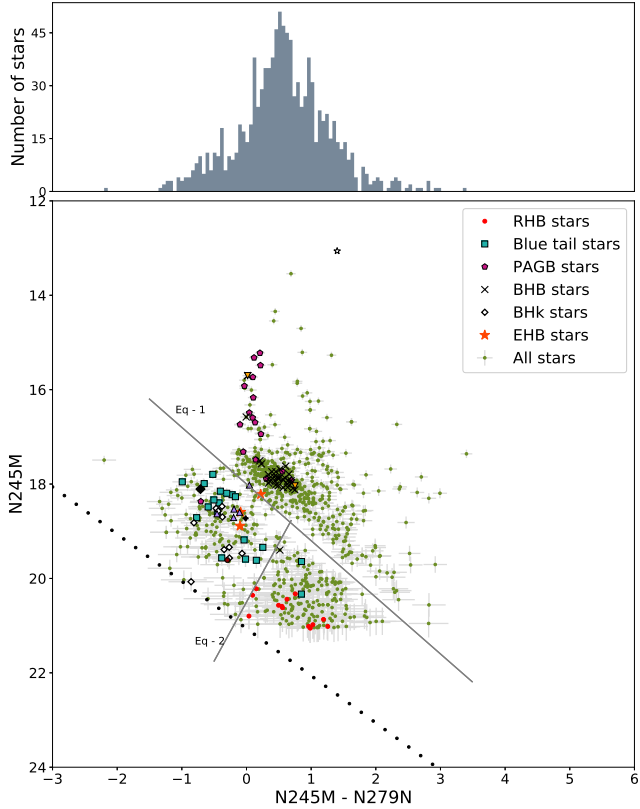}
\caption{The color-magnitude diagram constructed using N245M - N279N versus N279N. The green filled circles represent stars common to N245M and N279N filters which have not been identified in earlier studies. The other symbols are the same as those shown in figure \ref{fig:fuvnuvcmd}. The solid lines are used to demarcate the gaps (discussed in the text) and are represented by Eqns.~\ref{eq:2451} and \ref{eq:2452}.} 
\label{fig:N245N279}
\end{figure}

In the NUV - NUV plane, we present two color-magnitude diagrams. We first select two NUV filters that probe distinct wavelength ranges with each filter exhibiting a bimodal distribution in magnitude, viz. N245M and N279N. The CMD of N245M - N279N versus N245M, constructed using these filters is shown in Fig.~\ref{fig:N245N279}. The CMD reveals three prominent groups and the cluster members identified from literature help in ascertaining the nature of stars in each group. A clear gap is observed to run diagonally across the CMD that segregates one group of stars from the other two. The distinct gap can be characterised by a straight line in the CMD plane, given by the following expression
\begin{equation}
\rm{N245M}=1.2\times(\rm{N245M}-\rm{N279N})+18.0
 \label{eq:2451}
\end{equation}
The filter names in the equations refer to magnitudes associated with the respective filter. To the upper right of the gap associated with Eqn. (1) lie the BHBs. The BHB group of stars are relatively bright in NUV and occupy the region on the brighter end of the magnitude axis but span a wide range on the color axis, -0.5 to 3. The sparse distribution of stars that are luminous represent the Post-AGB and Post-Early AGB stars. To the lower left of the gap are seen groups of stars belonging to the RHB, EHB and BHk stages. These sets of stars are divided into two groups with a gap that is not very broad. This gap can be represented by a straight line represented by the following equation:
\begin{equation}
\rm{N245M}=2.5\times(\rm{N279N}-\rm{N245M})+20.5
 \label{eq:2452}
\end{equation}
The RHB group occupies a region within the color range 0 to 2.5 while lying on the fainter end of the magnitude distribution in N245M filter. It is to be noted that this combination of filters helps in the identification of the RHB stars, which were not detected in FUV. The RHB stars are cooler and fainter as compared to hot BHB stars. As a result, they are very faint in FUV while they are relatively bright in NUV, enabling their detection. The second group of stars to the left of the gap occupying regions of brighter magnitude are the EHB and BHk stars. This group appears segregated with respect to the RHB group although the gap is not very broad. 


\begin{figure}
\includegraphics[width=1\linewidth]{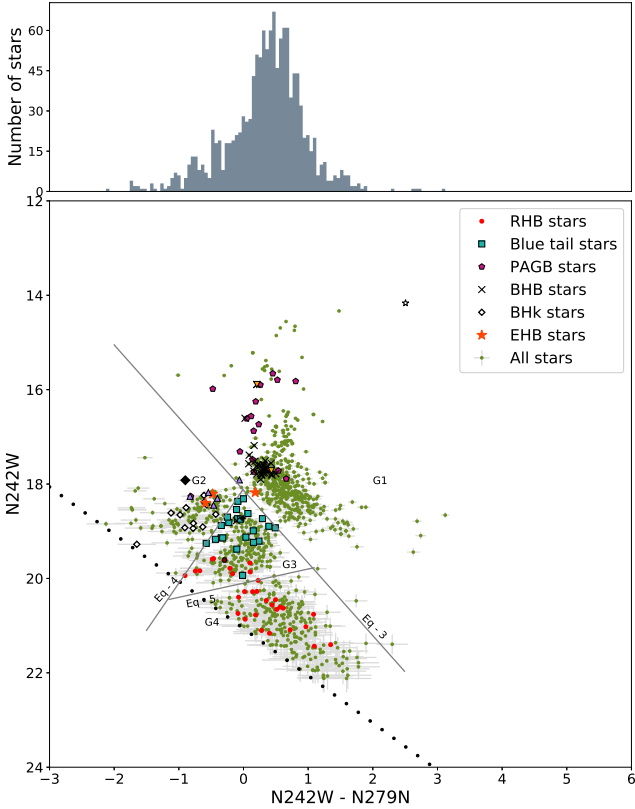}
\vskip -0.4cm
\caption{The color-magnitude diagram constructed using N242W - N279N versus N242W. The green filled circles represent stars common to N242W and N279N filters which have not been identified in earlier studies. The other symbols are the same as those shown in figure \ref{fig:fuvnuvcmd}.  The solid lines are used to demarcate the gaps (discussed in the text) and are represented by Eqns.~\ref{eq:2421}, \ref{eq:2422} and \ref{eq:2423}.}
\label{fig:nuvnuvcmd}
\end{figure}

An alternate combination of NUV filters that segregates the UV cluster members into four groups is the conjunction of N242W and N279N. These two NUV filters are not exclusive and the wavelength range of N279N filter lies within the broad band of N242W towards the red extremity. Both the filters show bimodal magnitude distributions, but the resultant color shows a unimodal distribution. This is apparent from the CMD constructed using N242W - N279N versus N242W, shown in Fig.~\ref{fig:nuvnuvcmd}. In this CMD arena, the filter combination enables a separation of cluster members into four groups. The four stellar groups are designated as G1, G2, G3 and G4. The gaps separating the four groups are represented by lines, given by the following equations:
\begin{equation}
\rm{N242W}=1.5\times(\rm{N242W}-\rm{N279N})+18.1
 \label{eq:2421}
\end{equation}
\begin{equation}
\rm{N242W}=2.0\times(\rm{M279N}-\rm{M242W})+18.1
\label{eq:2422}
\end{equation}
\begin{equation}
\rm{M242W}=0.3\times(\rm{N279N}-\rm{M242W})+20.1
\label{eq:2423}
\end{equation}
As earlier, based on the identification of cluster members from previous studies, we attempt the recognition of different stellar groups. G1 comprises of BHB, Post-AGB and Post-early AGB stars. The BHB stars occupy the fainter end of this group and the UV super-luminous stars appear as the brighter extension of this group. G2 consists of EHB and BHk stars. Unlike the other two groups, G3 and G4 show unique features in the sense that RHB stars are present in both the groups. Most of the known RHB stars are present in G4, while few are brighter and present in G3. These two groups are distinguished by a gap suggesting the segregation of RHB stars into two groups. 

\begin{figure}
\hskip -0.75cm
\includegraphics[scale=0.66]{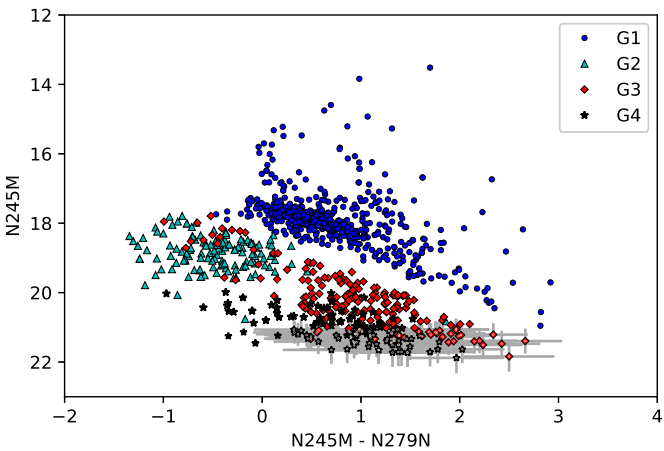}
\vskip -0.3cm
\caption{The color-magnitude diagram constructed using N245M - N279N versus N245M. The colours represent different groups of stars identified from Fig.~\ref{fig:nuvnuvcmd}. The open symbols with errorbars (shown in grey) represent stars below the sensitivity limit considered, i.e. stars identified in these two filters (N245W and N79N) whose $2.5\lesssim SNR \lesssim 3.5$. }
\label{fig:245_242grps}
\end{figure}

In order to validate this segregation, we located the positions of these groups of RHB stars in the NUV-NUV CMD considered earlier, i.e. N245M - N279N versus N245M (Fig \ref{fig:N245N279}). The various Groups of stars in the CMD are distinguished by different colors in Fig. \ref{fig:245_242grps}. The two groups of RHB stars are clearly differentiated in this CMD although no gap is present between the two groups. The G4 stars are bluer than the RHB stars of G3 in the N245M - N279N color and the merging of the two classes appears gradual. This visibly demonstrates the presence of two groups of RHB stars in this cluster. We notice that few G4 stars also belong to the EHB - BHk categories of stars. We speculated on the nature of these stars and found that these stars are FUV bright. This group of stars is unlikely to be affiliated to the RHB group since they are hot enough to be observed in FUV filters. Their distribution in CMD constructed using filters F154W and N279N indicates that these belong to the hot HB group. We term these stars as blue tail stars, as their nature (EHB/BHk member) is not certain. 

We distinguish between the two RHB groups (excluding the blue tail members), by naming the RHB stars in G3 as RHBI and those in G4 as RHBII. We note that few RHB members detected in the N242W and N279N filters are not detected above the considered sensitivity limits of N245M. However, they are faint and have been detected, albeit with a lower signal-to-noise ratio (SNR), $2.5\lesssim SNR\lesssim 3.5$. These have also been plotted with their respective errorbars in Fig.~\ref{fig:245_242grps} to increase the sample size of these RHB groups in the CMD. We find that the broad progression of RHBI and RHBII towards red color continues to hold. Our Group classification based on detection in filters N242W and N279N leads to identification of 621 stars in G1, 143 stars in G2, 247 stars in G3 (24 Blue tail stars and 223 RHBI stars), and 267 stars in G4.

\subsection{GAIA Color-Magnitude Diagram of UVIT Counterparts}
\begin{figure}
\hskip -0.75cm
\includegraphics[width=1\linewidth]{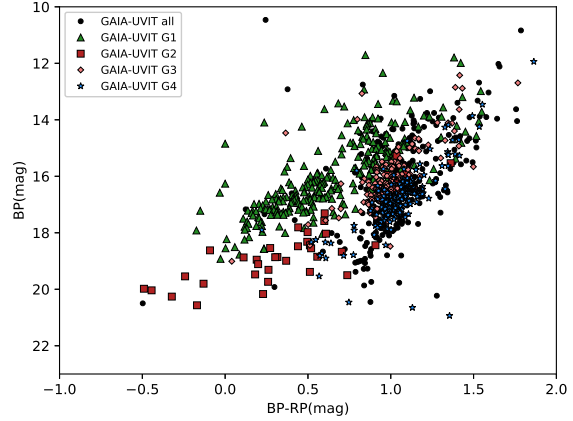}
\vskip -0.3cm
\caption{The color-magnitude diagram constructed using BP - RP versus BP passbands of GAIA showing stars detected with UVIT. The stars identified as G1, G2, G3 and G4 represent different groups of stars identified from Fig.~\ref{fig:nuvnuvcmd}. The black filled circles represent the other stars detected with the UVIT. }
\label{fig:gaia_cmd}
\end{figure}

We plot the hot UVIT stars in an optical colour-magnitude diagram to distinguish their location. We use the GAIA-DR2 catalog \citep{2018A&A...616A...1G} to identify and cross-match the UVIT sources. For a search radius of $1''$, we find that 
1267 UVIT stars have GAIA counterparts.  The photometric magnitudes of these stars through the GAIA Blue-Photometer (BP: 330 - 680~nm) and Red-Photometer (RP: 640 - 1000~nm) passbands \citep{2016A&A...595A...1G} have been used to create a colour-magnitude diagram (BP - RP versus BP) as the wavelength overlap is minimal. The GAIA colour-magnitude diagram is shown in Fig.~\ref{fig:gaia_cmd}. A gap is evident in the BP - RP colour cooresponding to $\sim7.5$~mag. The counterparts of UVIT stars belonging to various groups: G1, G2, G3 and G4 are marked with different symbols. We notice a conspicuous trend among the UVIT groups. Most of the G1 stars are concentrated towards the blue bank of the gap and there are few which lie on the brighter side of the BP magnitude range. This is in accordance with expectation as G1 comprises of BHB as well as the bright Post AGB and Post-early AGB stars. G2, comprising of EHB and BHk stars lie towards the fainter end of the blue branch branch. G3 and G4, that comprise of RHB stars are congregated towards the redder precincts of the CMD. G3 and G4 stars occupy distinct regions in the CMD although the transition from G3 to G4 is progressive in terms of brightness. They occupy the same range in colour: 0.8 - 1.2. As the overlap between the BP and RP filters is small ($\sim40$~nm), the colour can be used as an approximate temperature proxy. The notion that the two RHB groups have similar temperature ranges corroborates well with our findings from the UVIT colour-magnitude diagram of N245M - N279N versus N245M (Fig.~\ref{fig:245_242grps}). 

\subsection{Radial Density Distribution of UVIT cluster members}
\begin{figure}
\hskip -0.7cm
\includegraphics[scale=0.6]{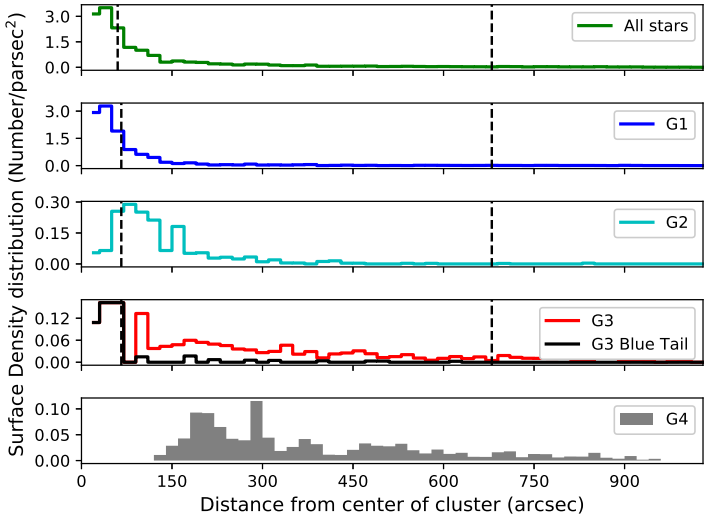}
\vskip -0.2cm
\caption{Histogram representing the radial distribution of the stellar surface density in steps of 20$''$. The regions in the top panel window displays the surface density distribution of all the stars that are detected through UVIT. The 
other panels represent the distribution of stars belonging to G1, G2, G3 and G4 respectively. In the window displaying G3, red indicates all the G3 stars while the black line outlines distribution of stars that belong to G3 but are also observed in FUV filters. The distribution plot of the detected G4 stars is shown in grey as the sample is incomplete towards the central regions.}
\label{fig:spatial_242}
\end{figure}

Having discerned the hot cluster members based on their location in the CMDs, we proceed towards  inspecting the relative distributions of various stellar groups across the cluster. In order to accomplish this, we ascertain the radial distance of each star from the cluster centre, taken as ($\alpha$, $\delta$)$_{\rm J2000}$ = $9^h 12^m 03.10^s$, $-64^\circ 51'48.6''$ and estimate the surface density of stars in annular regions about the centre. The histogram displaying the radial surface density distribution of all stars in the cluster is shown in the top panel of  Fig.~\ref{fig:spatial_242}. Stars have been detected upto an outer radius of $17'.16$. 
Since the pointing of UVIT during our observations of NGC 2808 was not centered at the cluster centre, we consider a symmetric region about the centre, which corresponds to a radius of 10$'$.82. We note that the tidal radius of the cluster is 15$'$.55 \citep{2000A&A...363..159B}. The radial region within the two vertical dashed lines (corresponding to $60''$ and $10'.82$ , respectively) represents the region where the stars extracted are believed to be largely complete. We have tested the completeness of our sample by adding artificial stars of varying intensity to the images and extracted them using the methods described earlier. For instance, we find that the extraction is 100\% complete in N242W for stars upto extinction corrected magnitude of 20.2~mag from the central $60''$ outwards. We find that this completeness magnitude represents the brighter edge of G4 group of stars (see Fig.~\ref{fig:nuvnuvcmd}). It is to be noted that 100\% completeness is achieved at fainter magnitudes in the outer regions as the cluster spatial density decreases. The radial profile plot of G4 is therefore for a sample that is incomplete in the inner regions. 

The surface radial density distributions of the total number of stars as well as various Groups of stars is displayed in Fig.~\ref{fig:spatial_242}. The radial density distribution of total number of stars is seen to decline with distance from the cluster centre. We observe that G1 or the number of post-AGB as well as BHB stars falls off with radial distance monotonically. The stars belonging to G2 category, i.e. EHB and BHk display a flattened distribution till $\sim100''$ and the number declines thereafter. This is in contrast to the radial distributions of G3 and G4 that appear to be nearly uniformly distributed across the cluster. G3 comprises RHBI  stars as well as blue tail starss, while G4 consists of RHBII stars. We observe that in G3, the blue-tail stars group close to the centre. Both RHBI and RHBII stars are found to extend to outer radial regions till the cluster tidal radius.

\section{Discussion}
\label{sec:discussion}
Gaps and jumps have been demonstrated to be ubiquitous features in the CMDs and color-index diagrams (CIDs) of HB populations in globular clusters. These are believed to be intricately linked to inherent processes in stellar interiors and atmospheres \citep{2016ApJ...822...44B}. We have observed this cluster anew in the near and far-ultraviolet space using UVIT filters spanning across the wavelength range $130 - 300$~nm. 

\subsection{Bimodality in the NUV Magnitude distribution}

We observe a bimodality in the magnitude distributions of stars detected in most of the NUV filters. We investigate the nature of the bimodality observed in different filters based on our classification of the cluster members into distinct Groups. It is to be noted that the group classification is applicable to the subpopulation ($\sim60\%$) that have been detected in both the filters N242W and N279N.
Fig.~\ref{fig:mag_hist} displays the contributions of G1 (BHB and PAGB), G2 (EHB and BHk), G3 (RHBI and blue tail) and G4 (RHBII) populations to the total number of stars in the magnitude distribution. For all the NUV filters, we find that the BHB or G1 stars always lie on the brighter end of magnitude range whereas the RHB stars from G4 and G3 constitute the red edge of the histograms. 

We perceive that the contributions of various group of stars giving rise to bimodality is different in the UV wavebands (filters) under consideration. In the broad band N242W filter, the bimodality essentially arises due to differences in luminosities of RHBI and RHBII, leading to a gap that defines the bimodality. The blue tail population of G3 has an overlap with the G2 population. The luminosities of EHB and BHk stars (G2 population) is intermediate to the RHB and BHB stars. The latter constitute the brighter extremity of the primary peak. In N245M and N263M, we notice that the bimodality appears to be due to the contrast in luminosities between the RHB and hot HB populations. In N279N, the narrow primary peak is composed of hot BHB stars (i.e. G1), while the broad secondary peak has contributions from the remaining populations. Another striking feature is the shift of magnitude distribution of G2 stars from N219M to N279N filters. The median of G2 magnitude distribution becomes fainter (by $\sim0.8$~mag) from N219M to N279N filter, i.e. as wavelength increases. Considering that they represent the hottest HB populations, this is in accordance with expectation. The RHB group of stars are barely detected in N219M filter. This particular waveband effectively probes the shortest of the NUV wavelengths. Hence we attribute the relative lack of detection of RHB stars to their lower photospheric temperatures.

 The magnitude corresponding to the minimum between the peaks of the bimodal distribution also shifts with wavelength: 19.8~mag in N245M, 19.1~mag in N263M and 18.3~mag in N279N. In other words, the specific flux density establishing the bimodality becomes larger with increasing wavelength. As the flux contribution by specific groups to a given filter is a reflection of the underlying stellar atmospheric processes, detailed stellar evolution modeling is vital for understanding the nature of the bimodality and the specific flux density defining the bimodality. The evolution of  bimodality with wavelength observed for NGC 2808
demonstrates the power of UVIT's multiple filters to advance our understanding of stellar atmospheric processes of the RHB group.

\subsection{Gaps in UVIT CMDs}

The UV magnitude distributions also manifest as gaps and accumulations of stars in specific regions of CMDs. In F154W - N279N CMD, the gap that separates the BHB from the EHB and BHk stars can be attributed to the bimodality in the magnitude distribution of N279N filter. This specific FUV - NUV CMD can be compared to that constructed by \citet{2001ApJ...562..368B} for the same cluster using filters similar to the ones used in the current work: HST STIS FUV/F25QTZ and NUV/F25CN270 filters for the same cluster. It is to be noted that a direct comparison of CMDs is difficult in the absence of colour equations connecting the two photometric systems with different filter characteristics. 
Consequently, the gap separating the BHB stars from the EHB and BHks stars are perceived at different values of color and magnitude.

The BHB population provides a tighter clustering unlike the EHB and BHk populations of stars (see Fig \ref{fig:nuvnuvcmd}). We note that the blue tail stars  appear to belong to the group of EHBs and lie close to the gap. However, they are different from EHBs and BHks, as is evident from the N242W - N279N CMD. In addition, we observe from the FUV - NUV CMD that there are a group of objects located in the region where ZAMS track is expected. These are encircled in Fig.~\ref{fig:fuvnuvcmd} and we suspect them to be BS candidates.  These stars are expected to lie at the fainter end of the extension of zero age HB loci, as demonstrated by \citet{2001ApJ...562..368B}. 

The gap separating the BHB from the EHBs is the  M - jump \citep{2004A&A...420..605M}, corresponding to T $\sim$ 20,000~K. This gap is prevalent in numerous globular clusters \citep{2016ApJ...822...44B}. This M - jump appears between G1 and G2 stars in the N245M - N279N CMD (Fig \ref{fig:N245N279}). This CMD shows a broad starless gap separating the BHBs (G1) from other groups of stars that includes EHBs, BHks and RHBs. The progression from RHBs (G3 and G4) to G2 is gradual and a tentative paucity of stars is seen between the RHB group and G2. The same M - jump that separates BHBs from EHBs and BHks, also extends further and segrategates the BHBs and RHBs. This would suggest a similar underlying mechanism for separation between both the groups. Thus, the UV CMDs enable us to analyse the gaps as a common feature unlike the optical and UV-optical CMDs \citep{1998ApJ...500..311F} and CIDs which divulge the gaps as discrete, i.e. in distinct locations. 

The two RHB groups of stars are set apart by a marginal gap (described by Eqn. 5) in the N242W - N279N CMD. This gap is not apparent in CID of this cluster described by \citet{2016ApJ...822...44B}. It is likely that this is analogous to the two RHB groups observed in other metal-rich globular clusters such as NGC 6637, NGC 6352, NGC 5927. While there is a large body of work on the extended HB morphology of NGC~2808, we could not find any work that substantiates the presence of two RHB groups of stars in this cluster. There is, however, the tantalising  spectroscopic evidence provided by \cite{2014MNRAS.437.1609M} based on Na abundances of 96 HB stars that indicated that there could be two groups of RHB stars. Based on the Na content, they observed a bimodality which is not consistent with a single RHB population. They also observed a gradient in colors of these RHB stars but the uncertainties did not allow the confirmation about the nature of star-formation histories. 
With our N242W - N279N diagram, we distinctly perceive two groups of RHB stars. It is difficult to gauge their temperatures as the progression is gradual and both groups occupy similar temperature ranges evident from the colour-magnitude diagrams of N245M - N279W and BP - RP. It is likely that this is responsible for the lack of distinct bimodality in Na-abundance. We also observe that the G3 stars are brighter compared to the G4 stars. Thus, our photometric method enables the segregation of 490 RHB stars into the two groups.

\subsection{Multiple UV populations in view of evolutionary scenarios}

We next speculate on the origin of the RHB groups of stars. The distinct RHB groups in UV CMDs appear to point towards distinct star formation histories. The main-sequence (MS) of this GC is comprised of quintuplet stellar populations, labeled A, B, C, D and E by \citet{2015ApJ...808...51M}. The mid-blue and extreme-blue MS (D and E, respectively), associated with higher helium enrichments, are believed to give rise to the BHB and EHB stars. The red MS (rMS) comprises of three branches (A, B and C) and we anticipate that the two RHB groups are plausibly related to the rMS branches discerned. 
The rMS branches are likely to have evolved to distinct RHB populations.  However, which rMS groups and/or their combinations are responsible for RHBI and RHBII is not known with the information available and detailed investigation is required to ascertain this. 

In case of NGC~2808, different stellar populations show distinct spatial distributions. While the hotter G1 and G2 are concentrated in the core, the cooler RHB stars (G3 and G4) have a distribution extending outwards. 
Earlier studies investigating the radial distribution of HB stars \citep{{2009ApJ...696L.120I},{2000A&A...363..159B}} did not find any significant radial trend from centre to outward regions. However at $1.5\sigma$ level, \citet{2009ApJ...696L.120I} did find a trend, i.e. a deficiency in RHB stars towards the centre (radius $\lesssim 35''$). 
An alternate study by \citet{2016MNRAS.463..449S} compares the radial distributions of the multiple populations of MS stars (i.e. A, B, C, D, E). These authors find that the D and E stars are concentrated towards the centre while the A+B+C stars  extend outwards with a progressive scarcity towards the central region (they investigated annular regions between $45''$ and $9'$). As the D and E are believed to be the pre-cursors of G1 and G2 helium enchanced second generation of stars, these results are in agreement with our observations. This is because according to multiple population formation models  \citep{2008MNRAS.390..693D}, intermediate and blue MS would give rise to BHB and EHB stars. The scarcity of A+B+C stars towards the centre observed by \citet{2016MNRAS.463..449S} is difficult to directly corroborate with the spatial distribution of RHB stars from this work. However, the fact these groups of stars are extended to large radial distance suggests that the RHB groups of stars could represent the evolved populations of the rMS.

The spatial distribution of cluster members from the present study can be used to comment on the evolutionary scenarios proposed for the formation of multiple populations. We consider the AGB ejecta scenario where the formation of multiple generations of stars in globular clusters has been investigated through simulations \citep{{2008MNRAS.391..825D},{2008MNRAS.390..693D}}. In this framework, the gas near the centre is cleared by the supernova explosions of first generation stars. A cooling flow develops as a result of which the CNO processed winds from massive AGBs are retained by the cluster potential as they are slow. This provides the gas necessary for the formation of second generation stars. The first generation stars, associated with RHB, have normal helium abundance while the later generation BHB, EHB and BHk stars have enhanced helium abundances as they are formed from processed gas. Thus, it is expected that the later generation stars would be concentrated within the inner core of a more extended first generation population. 

\begin{figure}
\includegraphics[width=1\linewidth]{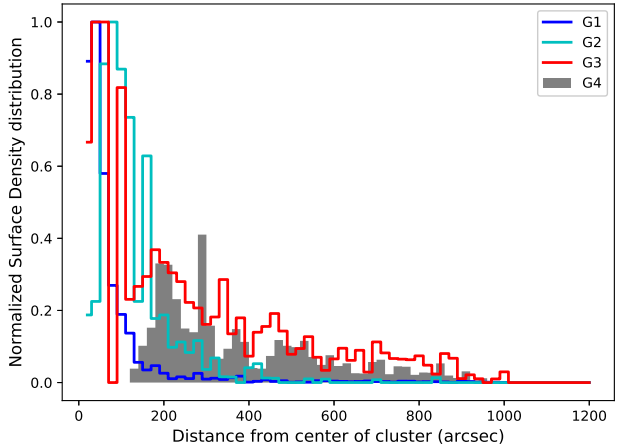}
\caption{Histograms of the radial stellar surface density, normalised at the value corresponding to radial bin of $60''$ for G1, G2 and G3. The distribution of G4 is normalised at the value of G3 corresponding to radial distance of $200''$.}
\label{fig:norm_density}
\end{figure}

More recently, \cite{2016MNRAS.458.2122D} have attempted to explain the five generations of stars using the same model. In addition to AGB ejecta, they also invoke the accretion of pristine gas to explain the multiple populations. They arrive at the BEDCA scenario for the main sequences A, B, C, D and E. According to this scenario, the rMS type B corresponds to the first generation, while C and A types of rMS correspond to the last generation. The mid- and extreme-blue MS (D and E) correspond to the stars of the second and third generations, respectively. From our spatial distribution, we find that the hot HB population (G1 and G2) are abundant near the centre which is consistent with the premise that they are related to the D and E types. The G1 population which corresponds to MS type D is more centrally concentrated than G2, associated with EHBs and BHks (MS type E). The normalised spatial density distribution is shown in Fig.~\ref{fig:norm_density} for comparison. This is in excellent agreement with the suggestion that G1 stars which are more centrally concentrated formed later than G2 with a broader distribution. 

The spatial distribution of  the RHB classes are similar and show broad radial distributions compared to G1 and G2. If the RHB groups of stars correspond to evolution associated with the A and/or C rMS, then it is expected that few of these stars would be more concentrated towards the centre than G1 and G2, rather than distributed in the outer radial regions. It is possible that G3 RHBI stars are related to B rMS. Whether G4 type RHBII stars are associated with earliest B rMS or later AC rMS is difficult to authenticate due to lack of detected stars near the central region. The spatial density of RHBI and RHBII stars in the outer radial regions are higher compared to G1 and G2 stars. If the rMS are believed to be the precursors to the RHB (as suggested from similar spatial distributions of rMS and RHB stars), then it would appear that all groups of rMS formed within short times of each other, unlike what has been proposed in the BEDCA model. With the current RHB identification, it should be possible to carry out detailed spectroscopic studies and ascertain abundances that will shed light on the formation scenario in greater detail.  

\section{Conclusions}
\label{sec:con}
The photometric results for the Galactic globular cluster NGC 2808 from an ultraviolet study using multiple filter of UVIT, shows that the cluster is an abode to exotic stellar population that includes EHB and BHk stars apart from the classic HB stars (BHB and RHB). We have identified for the first time the split in RHB population photometrically. This is another supporting evidence for multiple stellar population that resides in the GC.  Based on CMDs constructed using multiple FUV and NUV filters, we conclude that:

\begin{enumerate}
\item There is a bimodal distribution of stars on FUV-NUV CMD, which divides the group of stars into BHB and hot blue stars such as EHB and BHk stars.
\item A third group of stars is also seen in N245M - N279N CMD which comprises of relatively cooler RHB stars.
\item In N242W - N279N CMD, the RHB of the cluster segregates into two subgroups (RHBI and RHBII).
\item Based on the gaps seen in N242W - N279N CMD, we have divided the stars into four groups, G1 are the BHB and PAGB, G2 comprises the EHB and BHk stars, G3 has mixed population of blue tail stars and RHBI stars, and G4 consists of cool RHBII stars.
\end{enumerate}

The presence of stellar groups that may belong to various generations of stellar population is backed by the spatial extent of various group of stars and conforms broadly with the suggested evolutionary scenarios.  The hotter stars are concentrated towards the center of the cluster, while the cool stars are spread uniformly across the field of view away from the cluster center. There is a need to understand the formation as well as spatial distribution of the RHB groups of stars in the cluster as it does not comply with the current AGB ejecta models of formation of RHBs. The current study provides new information and insights regarding
  NGC 2808, which can be useful to understand the genesis of globular clusters.

\vspace*{1cm}

\noindent \textbf{Acknowledgements} \\

We thank the referee for providing useful comments that have helped in the improvement of the paper. This publication uses data from the AstroSat mission of the Indian Space Research Organisation (ISRO), archived at the Indian Space Science Data Centre (ISSDC).

\bibliographystyle{mnras}
\bibliography{ref_uvit} 

\bsp	
\label{lastpage}
\end{document}